\documentclass[12pt,letterpaper]{article}

\usepackage{graphicx}                       
\usepackage{geometry}                       
\usepackage{amsmath,amsfonts,bm}            
\usepackage{enumitem}                       
\usepackage[authoryear]{natbib}                        
\usepackage[hidelinks, colorlinks=true, 
linkcolor=black, citecolor=blue]{hyperref}  
\usepackage{fancyhdr, calc, lastpage}       
\usepackage{booktabs}                       
\usepackage{lipsum}                         
\usepackage{endnotes}                       
\usepackage{endfloat} 

\fancyheadoffset[L,R]{\marginparsep+1cm} 
\fancyfootoffset[L]{\marginparsep+1cm} 
\newcommand\setheader[2]{
    \fancyhead[L]{\footnotesize #1}
    \fancyhead[R]{\footnotesize #2, JASA-EL}
}   

\renewcommand\title[1]{{\linespread{1} \noindent\LARGE \bf \hskip2.25pc \parbox{.8\textwidth}{%
\LARGE \bf \begin{center} #1 \end{center}\rm } \rm\normalfont\normalsize} }
\renewcommand\author[1]{{\linespread{1} \noindent\hskip2.25pc \parbox{.8\textwidth}{%
   \normalsize \bf \begin{center} #1 \end{center}\rm } \vskip-1.4pc }}
\newcommand\address[1]{{\linespread{1} \noindent\hskip2.25pc \parbox{.8\textwidth}{%
   \footnotesize \it \begin{center} #1 \end{center}\rm }  \normalsize \vskip-1pc }}
\newcommand\email[1]{\vskip-.3cm \noindent\parskip0pc\hskip2.25pc \footnotesize%
   \parbox{.8\textwidth}{\begin{center}\it #1 \rm \end{center} } \normalsize  \vskip-.2cm}
\newcommand\PACS[1]{\vskip-2.75pc \begin{center}\parbox{.8\textwidth}{\small\bf PACS numbers: \rm #1 \hfill} \end{center}\vskip4pt}%

\renewenvironment{abstract}
{\vskip1pc\noindent\begin{center} \begin{minipage}{.8\textwidth} {\bf Abstract: } }
{ \vspace{.25cm} \end{minipage}\end{center}\normalsize\vskip-1.5pc}%

\def\fps@table{h}
\renewcommand\refname{\normalsize References and links \rm}

\makeatletter
\newcommand\@MaxCapWidth{4.25in}
\long\def\@makecaption#1#2{%
  \small
  \vskip\abovecaptionskip
  \sbox\@tempboxa{#1. #2}%
  \ifdim \wd\@tempboxa >\@MaxCapWidth
    \hskip2.25pc\parbox{4.5in}{#1. #2}
  \else
    \global \@minipagefalse
    \hb@xt@\hsize{\hfil\box\@tempboxa\hfil}%
  \fi
  \vskip\belowcaptionskip\normalsize}
\makeatother

\makeatletter
\renewcommand\@seccntformat[1]{\csname the#1\endcsname.\hspace{.1cm}}

\renewcommand\section{\@startsection {section}{1}{0pt}%
                                     {-2ex plus -1ex minus -.2ex}%
                                     {0.65ex plus 1.2ex}%
                                     {\normalsize\bfseries}}
\renewcommand\subsection{\@startsection{subsection}{2}{0pt}%
                                     {-2.25ex plus -1ex minus -.2ex}%
                                     {.45ex plus .2ex}%
                                     {\normalsize\itshape}}
\renewcommand\subsubsection{\@startsection{subsubsection}{3}{0pt}%
                                     {-2.25ex plus -1ex minus -.2ex}%
                                     {1ex plus .2ex}%
                                     {\small\upshape}}
\makeatother

\makeatletter
\let\old@theendnotes\theendnotes
\renewcommand{\theendnotes}{\old@theendnotes\vspace{.3cm}}
\makeatother

\makeatletter
\renewenvironment{thebibliography}[1]
     {\section*{\refname}%
      \@mkboth{\MakeUppercase\refname}{\MakeUppercase\refname}%
        \footnotesize

        \ifnum\value{endnote} > 0
        \theendnotes 
        \fi

      \list{\@biblabel{\@arabic\c@enumiv}}%
           {\settowidth\labelwidth{\@biblabel{#1}}%
            \setlength\itemindent{0pt}
            \setlength\itemsep{-1pt}
            }}
     {\endlist}
    
    \expandafter\ifx\csname natexlab\endcsname\relax\fi
    \expandafter\ifx\csname url\endcsname\relax
    \def\url#1{\texttt{#1}}\fi
    \expandafter\ifx\csname urlprefix\endcsname\relax\fi
    
    \providecommand{\noopsort}[1]{}
    
\makeatother

\begin{document}

\setheader{Unofficial template}{Lagrange}

\title{The bag-of-frames approach:\\ a not so sufficient model for\\ urban soundscapes} 

\author{Mathieu Lagrange (\texttt{mathieu.lagrange@cnrs.fr})  \\  IRCCYN, Ecole Centrale de Nantes, France \\
Gr\'egoire Lafay  (\texttt{gregoire.lafay@irccyn.ec-nantes.fr})  \\  IRCCYN, Ecole Centrale de Nantes , France, \\
Boris Defreville (\texttt{boris.defreville@gmail.com})  \\  ORELIA, France  \\
Jean-Julien Aucouturier  (\texttt{aucouturier@gmail.com}) \\  IRCAM STMS CNRS UPMC, France}
\address{}
\email{}

\begin{abstract}

The ``bag-of-frames" approach (BOF), which encodes audio signals as the long-term statistical distribution of short-term spectral features, is commonly regarded as an effective and sufficient way to represent environmental sound recordings (soundscapes). The present paper describes a conceptual replication of this approach using several soundscape datasets, with results strongly questioning the adequacy of the BOF approach for the task. As demonstrated in this paper, the good accuracy originally reported with BOF likely result from a particularly permissive dataset with low within-class variability. Soundscape modeling, therefore, may not be the closed case it was once thought to be.

\end{abstract}

\section{Introduction}

In this paper, we study a particularly common pattern-recognition paradigm, the ``bag-of-frames'' approach (BOF), which represents audio signals as the long-term statistical distribution of their short-term spectral features. Typically implemented as a Gaussian mixture model (GMM) of Mel-Frequency Cepstrum Coefficients (MFCCs), the BOF approach is commonly applied to model music \citep{SU14} and soundscapes \citep{Barchiesi2015} and is a standard component in several audio pattern recognition software packages \citep{PAG13}. Further, recent psychoacoustical evidence suggest the approach bears some resemblance with human auditory processing for sound textures \citep{McDermott2013, Nelken2013}. 

In an influential 2007 article, Aucouturier, Defreville \& Pachet \citep{Aucouturier2007} applied a BOF model to categorize both polyphonic music and soundscapes. Their results showed that, while BOF was a meriting model for their polyphonic music dataset, it was spectacularly effective for soundscapes, reaching accuracies of 96\%. The contrast, they interpreted, lied in differences in the temporal structure of both types of stimuli, with music being more formally organized and soundscapes more easily summarized by statistics. In a later companion study \citep{AUC09}, they showed that soundscapes could be time-shuffled without altering listeners' perception of their acoustic similarity, while music could not. While more work was needed for music, the authors therefore concluded that BOF was a sufficient model to approximate human perception for soundscapes, practically ruling out the need to recognize the local acoustic events in a texture in order to identify it. 

That article, which is commonly regarded as the birth act of BOF modeling for audio, was influential in the soundscape community\footnote{If any indication, the number of citations for the paper as we wrote was 127 (source: Google Scholar, retrieved June 2015), 60 (47\%) of which included the words ``soundscape'' or ``environmental sound''}. BOF found applications in audio sound scene recognition \citep{Barchiesi2015}, audio event identification \citep{Giannoulis2013b} and soundscape quality assessment \citep{PAR14}. Because of its original comparison with more-demanding music stimuli, soundscape BOF modeling was also taken as a benchmark against which new music models were developped, and standardized soundscape datasets were assembled to conduct algorithm evaluations \citep{Giannoulis2013b}. However, to the best of our knowledge, the adequacy of the BOF approach for soundscapes, mostly derived from a single seminal study, was never challenged. Not only did the original BOF article introduced a new task (soundscape modeling) to the world of pattern recognition, it also appeared to close the case in the same effort. 

In this paper, we conduct a direct replication of the 2007 BOF article, using not only the original dataset, but three others, more recent soundscape datasets with less within-class variability. On these four datasets, we compare the original BOF model proposed by Aucouturier, Defreville and Pachet (2007) to the simple one-point averaging of sound features, on a standard task of categorization. We show that, while the excellent soundscape accuracy of BOF is replicated on the original dataset, it does not hold, by far, for more varied and realistic data. More surprisingly perhaps, for all datasets, BOF did not perform better than the much simpler one-point average approach. In other words, soundscape modeling may not be not the closed case it was once thought to be.

\section{Experimental Protocol}

\subsection{Algorithms}

We implemented the same BOF algorithm that was originally proposed in \citep{Aucouturier2007}, namely Gaussian Mixture Models (GMMs) of Mel-Frequency Cepstrum Coefficients (MFCCs). MFCCs are computed as coefficients from the Fourier transform along frequency of the signal's log spectrogram. By extracting MFCCs on every successive short-time window of a signal, it is transformed into a time-series of feature points in an euclidean space. Here, we extracted d=20 MFCC coefficients from 2048-point time windows using the AUDITORY toolbox implementation \citep{SLA98}.

GMMs are statistical models used to estimate a probability distribution $\mathcal{P}(x)$ as the weighted sum of $M$ gaussian distributions $\mathcal{N}_i, \forall{i<M}$, each parameterized by a mean $\mu_i$ and covariance matrix $\Sigma_i$, 
\begin{equation}
\mathcal{P}(x)=\sum_i^M{\pi_i\mathcal{N}_i(x,\mu_i,\Sigma_i)}
\end{equation}
where $\pi_i$ is the weight of gaussian distribution $\mathcal{N}_i$. Given a collection of MFCC feature vectors, viewed as samples from a random variable, the parameters $\pi_i$,$\mu_i$,$\Sigma_i, \forall i < M$ of a GMM that maximizes the likelihood of the data can be estimated by the Expectation Maximization (EM) algorithm. Here, we took M=50. 

In order to compare two audio signals $p$ and $q$, we estimate the parameters of a GMM for each signal's collection of MFCC vectors $p[n]$ and $q[m]$, and then compare the two GMMs $\mathcal{P}_p$ and $\mathcal{P}_q$ using their Kullback Leibler (KL) divergence: 
\begin{equation}
d_{KL}(p,q)=\int{\mathcal{P}_p(x)\log{\frac{\mathcal{P}_q(x)}{\mathcal{P}_p(x)}}}
\end{equation}. 

As a comparison for BOF, we also evaluated a simpler matching algorithm in which a signal's MFCC series $p[n]$ is averaged along time to a single mean MFCC vector $\tilde{p}$ of size d=20: 
\begin{equation}
\tilde{p}=\frac{1}{N_T}\sum_{t=1}^{t=N_T}{p[t]}, 
\end{equation}
where $N_T$ is the number of measured time points in the original representation. Two  average MFCC vectors can then be compared using the simple euclidean distance, defined as 
\begin{equation}
d_\epsilon(\tilde{p},\tilde{q})=\sqrt{\sum_i (\tilde{p_i}-\tilde{q_i})^2} 
\end{equation}
where $\tilde{p_i}$ and $\tilde{q_i}$ are the $i^{th}$ coordinate of average vectors $\tilde{p}$ and $\tilde{q}$. 

\subsection{Datasets} \label{sec:datasets}

\begin{table} [t]
\begin{center}
  
\caption{Morphology of the soundscape datasets used in this study. Where available, we also report categorization performance of humans (\%). \label{undu180}} 
\begin{tabular}{lccc|ccc} 
& \multicolumn{3}{c}{recordings} & \multicolumn{3}{c}{human} \\

dataBase & number & number per class & duration & p@5 & map & accuracy \\ 
\hline 
\texttt{AucoDefr07} &  16 &  4$\pm$1 & 947$\pm$597 & - & - & - \\ 
\texttt{Guastavino} &  16 &  3$\pm$0 &    15$\pm$1 & 96 & 99 & -  \\ 
\texttt{Tardieu} &  66 & 11$\pm$4 &    17$\pm$2 & 72 & 63  & - \\ 
\texttt{QMUL} & 100 & 10$\pm$0 &    30$\pm$0 & - & - & 75  \\ 
\end{tabular} 
\end{center} 
\end{table} 

We collected four datasets from the previous work of different laboratories (Table \ref{undu180}):
\begin{enumerate}
\item \texttt{AucoDefr07}: Used in the original BOF article, this dataset was recorded by Boris Defreville in Paris (France) and consists of 16 recordings distributed in 4 classes: ``avenue", ``street", ``market" and ``park". The first class consists of 5 recordings of 3 avenues, the second of 3 recordings of 2 streets, the third of 5 recordings of 2 markets and the fourth of 3 recordings of 1 park. For the work of \citep{Aucouturier2007}, these recordings were segmented into 3-minute units, giving 78 units with about 20 units per class. We are publishing this previously unreleased dataset as part of this paper: \url{https://archive.org/details/defreville-Aucouturier_urbanDb}.
\item \texttt{Guastavino}: provided by Catherine Guastavino and recorded in Paris (France) as material for psychoacoustics experiments \citep{Guastavino2007a}. The dataset consists of 16 recordings of less than 3 minutes, distributed in 5 classes: ``vehicles", ``vehicles and pedestrians", ``cafes", ``markets" and ``parks". Each class consist of 3 recordings of 2 places, except for the fourth which consists of 3 recordings of 3 places.
\item \texttt{Tardieu}: recorded by Julien Tardieu in Paris (France) for the purpose of psychoacoustics experiments \citep{Tardieu2008}, and consists of 66 recordings in 6 different French train stations distributed in 6 classes: ``Platforms", ``Halls", ``Corridors", ``Waiting room", ``Ticket booths" and ``Shops". Classes are well balanced in terms of number of recordings and number of different locations.
\item \texttt{QMUL}: recorded in a wide variety of Greater London (UK) locations over Summer and Autumn 2012 by 3 members of Queen Mary University of London \citep{Giannoulis2013}. The dataset consists of 100 30-second recordings equally distributed among 10 classes: ``bus", ``busy street", ``office", ``open air market", ``park", ``quiet street", ``restaurant", ``supermarket", ``tube", and ``tube station". To ensure that no systematic variation in the recordings covaried with scene type, all recordings were made in moderate weather conditions, at varying times of day and week, and each operator recorded occurrences of every scene type. The public part of the dataset considered in this study is available on the C4DM Research Data Repository: \url{http://c4dm.eecs.qmul.ac.uk/rdr/handle/123456789/29}.
\end{enumerate}

These four datasets were not created equal in their potential to validate computational approaches. The former two appear limited in terms of variety of recording locations (4 and 5, resp.) and number of recordings per class (4 and 3, resp.). Results obtained on these datasets shall therefore be considered with care. The latter two, while still rather small for a classification task, boast a more realistic variety of recording locations (6 and 10, resp.) and number of recording per class (11 and 10, resp.). 


\subsection{Metrics}

We followed the same evaluation paradigm as \citep{Aucouturier2007}. Algorithms were evaluated on their ability to retrieve, for a given seed item, nearest neighbors that belong to the same category. This ability was measured with two standard metrics: 1) the precision at rank 5 (p@5), \textit{i.e.} the average number of items of the same class among the 5 closest items averaged for all possible seed items and 2) the Mean Average Precision (MAP), \textit{i.e.} the average precision across all possible ranks. For both metrics, the target rank is bounded by the total number of items in the category to which the seed item belongs. 

\section{Results}

We tested for the impact of a few algorithmic design variants within the general guidelines provided in \citep{Aucouturier2007}: audio recordings can be amplitude-normalized prior to encoding; the first MFCC coefficient (related to the energy at the frame level) can be discarded before modeling; models $A$ (learned on features $a$) and $B$ (learned on features $b$) can be compared using the Monte Carlo sampling (MCMC) method or by the direct marginalization on the features (i.e., by averaging the unconditional data probability of $a$ given model $B$ and the unconditional data probability of $b$ given model $A$  to estimate the distance between the models). We found that none of these factors greatly affected either metrics. In the following, we used the configuration that gave the best result on the \texttt{QMUL} dataset: audio normalization, first MFCC coefficient included, and MCMC sampling.

\begin{table} [t]
\begin{center} 
\caption{Categorization performance in terms of mean and standard deviation of the precision-at-rank-5 and mean-average-precision (P@5 / MAP) in percents.\label{results}} 
\begin{tabular}{lccc} 
dataBase & chance & Average & BOF \\ 
\hline 
AucoDefr07 & 44$\pm$20 / 37$\pm$8  & 96$\pm$13 / 82$\pm$17 & 97$\pm$12 / 83$\pm$16 \\ 
Guastavino & 28$\pm$10 / 58$\pm$10  & 39$\pm$14 / 67$\pm$17 & 40$\pm$10 / 69$\pm$14  \\ 
Tardieu & 33$\pm$18 / 33$\pm$7  & 49$\pm$22 / 43$\pm$17 & 48$\pm$20 / 42$\pm$13 \\ 
QMUL & 28$\pm$13 / 25$\pm$6  & 47$\pm$23 / 42$\pm$17 & 48$\pm$21 / 38$\pm$13 \\ 
\end{tabular} 
\end{center} 
\end{table} 

Table \ref{results} show categorization precision scores broken down by dataset. 
The evaluated computational approaches perform better than random, showing that some information is indeed retained by the models. Further, MAP scores do not strongly differ from p@5, suggesting coherent performance across different ranks (the large MAP values achieved with the \texttt{Guastavino} dataset are most probably due to its small size). 

While performance scores in our tests are consistent across the \texttt{Guastavino}, \texttt{Tardieu} and \texttt{QMUL} datasets, performance for the original \texttt{AucoDefr07} dataset clearly stands out. We believe that this result can be explained by two factors. First, in order to replicate results discussed in \citep{Aucouturier2007}, the recordings were divided into 3-minute units and these units were next categorized. This means that units that are close to one another will most probably come from the same recording. If this processing step is removed and each recording is categorized as a non-separable whole, we found precision dropped to 71 \%. Second, the average number of recording locations per class is only 2, which may have the same effect of biasing the results to favor recordings from identical locations. 

The average precision over the two most realistic datasets in this study, \texttt{Tardieu} and \texttt{QMUL}, was only 48 \%. We believe that this figure is a more appropriate benchmark of the adequacy of BOF for soundscape categorization than the 96\% obtained on the overly permissive \texttt{AucoDefr07} dataset. This figure can be compared to that obtained by BOF for musical genre classification (77\%) in \citep{Aucouturier2007}. 

Finally, for all four datasets, no statistical difference was found between the BOF approach and a simple one-point feature average. The computational complexity involved in computing and comparing GMM models therefore seems an unnecessary burden: not only is there no added advantage in considering several, rather than one, gaussian components per item, but there also does not seem to be any discriminative information encoded in the standard deviation, rather than in the sole mean value of the feature distribution. Furthermore, with simple averaging, recordings are mapped directly in a 20-dimensional Euclidean space, which allows faster matching and indexing methods than the GMM-based metric space.

\section{Discussion}

These new results challenge the commonly-held view that the BOF approach is a sufficient model to categorize soundscapes. While we replicated the excellent performance of BOF described in \citep{Aucouturier2007}, it appears that this figure likely results from an overly permissive dataset with abnormally low within-class variability: a majority of exemplars in each class were different extracts of the same recordings, with plausibly strongly similar acoustic characteristics that would not generalize across different recordings. For the more realistic (yet not overly complex) tasks associated with the three alternative datasets tested here, BOF performance was severely degraded (48\%). In fact, it did not score strongly better than a mere one-point average of the signal's features.

These results suggest that, while meaningful information for soundscape categorization is preserved in their summary statistics, this information by no means allows a sufficient level of generalization to handle categories with strong intrinsic variability. While recent results showed some improvements using more advanced statistical techniques than the basic BOF implementation tests here \citep{Barchiesi2015}, it remains doubtful whether purely textural models can solve the problem. 

Far from an anti-BOF epiphany, we believe these results should be taken to indicate that soundscape categorization is, and remains, a difficult computational problem when considered at realistic levels of within- and between-class variability. A more promising approach would likely combine textural modeling with prior recognition of individual sources. To this aim, efforts could be made at developing ecologically- and cognitively-grounded context models, linking events to textures (e.g. car horns, traffic sounds, pedestrian chatter to a ``busy street'' texture). Such efforts will need to confront such hard questions as what prior information can be reliably assumed about an event's distribution across soundscapes and about an event's acoustic signature, and what distinguishes an event from its background model. These are questions routinely and efficiently solved by the human auditory system \citep{BAL91,GUA05,LEE09,Nelken2013}; the present results show that our best computational models are not there yet. 
 
From a methodological point of view, the problem found here with the \texttt{AucoDefr07} dataset is reminiscent of a similar fallacy identified in the music information retrieval community for the task of musical genre classification. In early datasets contemporaneous with \texttt{AucoDefr07}, exemplars from a given genre category often consisted, by ease of construction, in several songs by the same artist or the same album - which artificially boosted the classification accuracy \citep{STU12}. Guidelines to avoid such biases were only recently introduced \citep{FLE07}, and debate is still ongoing as to what constitutes a dataset apt to evaluate the ability to recognize the concept of genre, rather than an incidental correlate thereof \citep{STU13}.  

More generally, this paper's critical re-examination of previously published results is one more argument in favor of direct and conceptual replications in science in general, and computational science in particular. This research was only made possible by our peers' willingness to share their research datasets or to publish them as part of evaluation campaigns \citep{Giannoulis2013}. These practices are making science a better place, and we were grateful to benefit from them here.     

\section*{Acknowledgments}

Research project partly funded by ANR-11-JS03-005-01(HOULE) and ERC Grant 335536 (CREAM). The authors would like to thank Catherine Guastavino, Julien Tardieu for sharing their datasets.


\end{document}